# Anticipation and the Non-linear Dynamics of Meaning-Processing in Social Systems

Loet Leydesdorff

Amsterdam School of Communications Research (ASCoR), University of Amsterdam,

Kloveniersburgwal 48, 1012 CX  Amsterdam, The Netherlands.

loet@leydesdorff.net ; http://www.leydesdorff.net

**Abstract**

Social order does not exist as a stable phenomenon, but can be considered as "an order of reproduced expectations." When anticipations operate upon one another, they can generate a non-linear dynamics which processes meaning. Although specific meanings can be stabilized, for example in social institutions, all meaning arises from a global horizon of possible meanings. Using Luhmann's (1984) social systems theory and Rosen's (1985) theory of anticipatory systems, I submit algorithms for modeling the non-linear dynamics of meaning in social systems. First, a self-referential system can use a model of itself for the anticipation. Under the condition of functional differentiation, the social system can be expected to entertain a set of models; each model can also contain a model of the other models. Two anticipatory mechanisms are then possible: a transversal one between the models, and a longitudinal one providing the system with a variety of meanings. A system containing two anticipatory mechanisms can become hyper-incursive. Without making decisions, however, a hyper-incursive system would be overloaded with uncertainty. Under this pressure, informed decisions tend to replace the "natural preferences" of agents and a knowledge-based order can increasingly be shaped.

# 1. Introduction

The social phenomenon of meaning can be modeled. Meaning can be understood as holding current value as well as an *anticipation* of possible futures. Rosen (1985) defined anticipatory systems as systems that contain a model of themselves. The modeling part advances on the modeled one and provides meaning to the latter from the perspective of hindsight. The time axis of the historical process can thus locally be inverted within a system (Dubois, 1998).

The modeling system can also be considered as a semantic domain in which historical developments can be appreciated and provided with value. From a biological perspective, Maturana (1978) distinguished between a first-order "consensual" domain in which organisms interact recursively, and a second-order one "within the confines of a consensual domain" in which organisms are additionally able to observe one another and provide one another's actions with interpretations. Such a domain would be "indistinguishable from a semantic domain" (*Ibid.*, p. 49). From this (biological) perspective, "language" would make it possible to recombine components of historically generated consensual behaviour into the generation of new consensual behaviour. Meaning would in this case be constituted by the exchange, and a next-order *social* system with its own dynamics would be shaped (Maturana, 2000).

Luhmann (1984) built his sociological theory on this next step by considering the *communication* of meaning as the operator of social systems: meaning is the medium of social systems and communication the operator. Meaning is selectively used and reproduced by communication. Thus, meaning is not merely generated and positioned as in a semantic domain, but can again be communicated using a dynamics which is proper to the social system and potentially different from those of individuals (Schutz, 1932; Berger & Luckman, 1966). The social system not only contains a model of itself, but the model can also be entertained within the system, for example, in a discourse (Leydesdorff, 2006).[1]

In biological systems, meaning is processed as wear and tear along the time axis in accordance with the "natural" (life-)cycle of the system. In social systems, specific meanings can be stabilized, for example in social institutions (Laland *et al*., 2000), but all meaning generated in inter-human communication arises with reference to a global horizon of possible meanings (Husserl, 1929). These cultural horizons of meaning can also change historically, as Weber noted in 1904:

> At one moment or another, the colour will change: the meaning of the perspective which was used without reflection will become uncertain; the road seems now to lead into zones of twilight. The light of the important problems of culture has gone beyond. (…) Science follows the constellations which make it a meaningful enterprise. (Weber, 1968, at p. 214).

Whereas this selection between the system and its environment is determined and therefore fixed in biological systems (e.g., by assuming natural selection), a non-linear dynamics of meaning-processing can be generated when this selection mechanism is considered as another degree of freedom in the communications between the system and its environment. Selections from global horizons of meaning add a third (globalizing) selection mechanism to momentary selections from historical events in the present and stabilizations of meaning over time.

---

[1] A reflexive individual can learn to handle these distinctions among different kinds of meaning (e.g., situational meaning, private meaning, etc.) and thus is able to develop communicative competencies (Leydesdorff, 2000).

When three subdynamics operate upon one another, all kinds of chaotic behaviour can be generated (Li & Yorke, 1975; May & Leonard, 1975). Thus, the emergence of social order can no longer be taken for granted, but needs always to be explained (Hobbes, 1651; Luhmann, 1995a). In other words, the social system can be expected to evolve as a non-trivial machine *because* it contains a non-linear dynamics of meaning (Baecker, 2002).

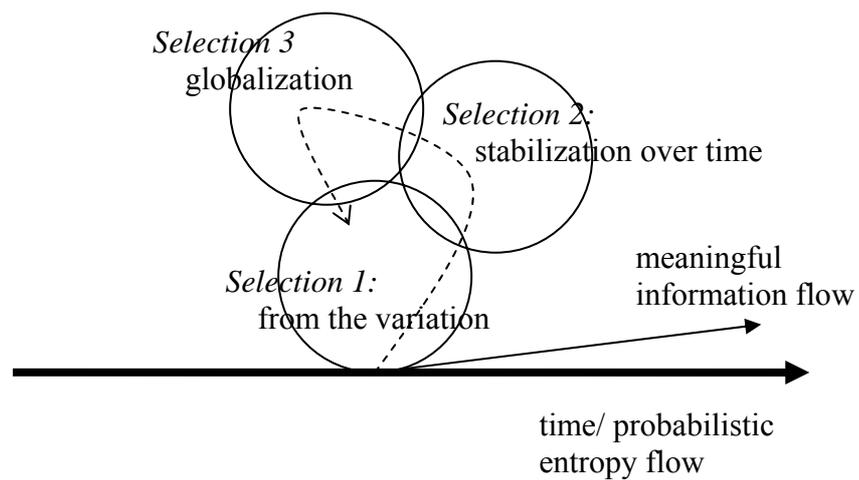

**Figure 1**: Three selections cycle in the case of processing meaning on top of the information flow.

The dashed arrow in Figure 1 indicates the resulting dynamics. The cycle may have a direction feeding backward or forward on the time axis, since the three selections can be expected to operate without *a priori* coordination. For example, selections in the present on the variation operate at the same time as selections from global horizons of meaning. However, one can expect that the results of the three subdynamics will be different, since selections operate asymmetrically.

Furthermore, the selections can operate upon each other: some selections can be mad for stabilization, while other are selected for globalization. Aelection mechanisms may thus co-evolve in a process of "mutual shaping." A third selection mechanism can be expected to disturb such mutual shaping by adding another source of uncertainty. In the case of biological systems, this third selection mechanism is determined as "natural" selection, and thus would drive the aging of the system. In Luhmann's social systems theory, the globalizing horizons of meaning tend to remain unspecified otherwise than with reference to Husserl's transcendental phenomenology (e.g., Luhmann, 1995b).

The three selection mechanisms have in common that meaning is provided from the perspective of hindsight and can therefore be modeled using incursive equations (Dubois, 1998; Leydesdorff, 2005). Incursive equations—to be explained below—invert the time axis by appreciating the past from the perspective of the present. Leydesdorff & Dubois (2004) have shown that the incursive formulation of the logistic equation can be used to model interaction and aggregation in inter-human communication. Interaction and aggregation can be considered as the building blocks of social systems of communication because (1) interaction is recursive (that is, interactions can operate on previous interactions), (2) interactive processes can be aggregated, and (3) aggregations can interact (Leydesdorff, 2003).

In this paper, I attempt to formulate how these building blocks can be organized into the architecture of a social system. How can the cycle of meaning processing—indicated in Figure 1—be closed as a feedback loop upon the linear movement of the entropy flux? First, I generalize the model—that is, the incursive formulation of the logistic equation—for the case of differentiated structures of meaning-processing at the supra-individual level. Secondly, I specify a hyper-incursive mechanism of anticipation as yet another possibility in a system which contains more than a single mechanism for anticipation. The hyper-incursive mechanism interacts with the incursive anticipations and makes historical decisions unavoidable. One can expect the resulting system to be *strongly* anticipatory (Dubois, 2000). Unlike weakly anticipatory systems—which entertain a model of themselves—strongly anticipatory ones cannot provide predictions because these systems use expectations for the construction of their future states (Collier, 2005).[2]

**2. Incursion**

A self-referential system can be considered as a function of itself and its environment. Baecker (2002, at p. 86) proposed modeling such a system using the following equation:

$$S = f(S, E) \qquad (1)$$

However, the environment is only relevant for a self-referential system as its carrying capacity; the environment sets limits to growth. This specific relation between the development of a system and its environment can be modeled using the logistic equation as follows (Devaney, 2003):

---

[2] Dubois (2003, at p. 114) defined hyper-incursion as an incursion that generates multiple future states.

$$x_{t+1} = ax_t(1-x_t) \qquad (2)$$

The carrying capacity of the selection environment $(1 - x_t)$ inhibits further growth of the system as a feedback term. For relatively small values of the parameter $a$ ($1 < a < 3$), this generates the well-known sigmoid curves of systems which grow and undergo transitions. For larger values of $a$, the model bifurcates into an oscillation (at $a \geq 3$) or increasingly generates chaos ($3.57 < a < 4$).

Unlike this biological (e.g., population-dynamic) model which operates with the arrow of time pointing forward, meaning is provided in the present with hindsight. This can be modeled using the same equation, but with different time parameters as follows (Dubois, 1998):

$$x_{t+1} = ax_t(1-x_{t+1}) \qquad (3)$$

In this case, the system builds on its previous state, but the current state provides the selection environment. Without any addition, Equation 3 can analytically be reorganized into Equation 4 as follows:

$$x_{t+1} = ax_t/(1+x_t) \qquad (4)$$

This system cannot generate chaotic phenomena because it converges into a stable state for all values of the parameter ($a$). Because incursion provides us with a model of a historically developing system, one is able to generate an observer in the simulation who reflexively follows the development of the observed system (Leydesdorff, 2005). The challenge, however, remains the construction of an observing system which is able to predict developments in the complex system under observation (Dubois, 2002).

**3. Longitudinal and transversal incursion**

All systems which entertain a model of themselves can provide meaning to the modeled system with hindsight, that is, by inverting the time axis locally. I shall call this the "longitudinal" generation of meaning, which will be distinguished below from the "transversal" generation of meaning. Transversal meaning can only be generated by systems which stabilize different codes for providing meaning. Thus, these systems can be expected to contain more than a single meaning. Furthermore, transversally generated meaning can again be provided with longitudinally generated meaning: the time axis stands perpendicular to the differentiation. In other words, the system spans a multi-dimensional space that develops over time.

Meanings may vary at each moment of time, but differently codified meanings can be stabilized only in systems that are *functionally* differentiated in terms of the codes of communication. The functionality of the differentiation means that the system is able to use the differentiation among the codes for its reproduction. By using the codes, events can then be appreciated differently within the system. Here, Luhmann's (1984) contribution becomes particularly valuable because he argued—following Parsons (1951)—that functional differentiation is prevalent in *modern* societies. He added that the differentiation should be considered as a differentiation of the symbolically generalized media of communication, that is, of next-order codification structures (Künzler, 1987).

Parsons (1967a and b) introduced "symbolic generalization" to explain the sometimes binding character of collective action. His concept remained close to Weber's concept of values, except that the values are not given; symbolic generalization would function as a mechanism to integrate the social system despite the structural differences within it (caused by functional differentiation). Using money as the model of an exchange medium, Parsons (1967a and b) extended this concept to include power and influence as other exchange media that can be generalized symbolically.

Luhmann (1975) elaborated on Parsons's concept by considering power, etc., as codes of communication, each of which can be generalized under the condition of functional differentiation. In a previous, hierarchical, order the different media were subsumed under each other: for example, the Holy Roman Emperor had to go to Canossa because the Pope had eventually the power to "excommunicate" him. Under the condition of modernity, exchange systems can develop according to their own logic, and thus science, politics, the economy, affection, etc., can further develop their specific codes of communication along (nearly) orthogonal axes. Symbolic generalization implies that every event can be assessed from the specific perspective of each code of communication. For example, everything can be assessed in terms of its economic value or its esthetic beauty, and these assessments do not have to correspond because they are coded differently.

Assessments using codes can be considered as events that can be assessed using a different perspectives. How do the assessments assess one another using different codes? For example, technologies which develop in generations over time have to compete for markets in the present. The markets code the technologies in terms of prices or price/performance relations, while engineers code the technologies according to professional standards. Thus, two codes of communication are operating simulataneously: the prices on markets and the professional coding expressed in engineering textbooks. The markets select in the present among the technologies which are developing over time. This can again be modeled using the incursive formulation of the logistic equation (Equation 3 above) because the development of the technology contains both a reference to a previous state of this technology (*i*) and a reference to a current state, but in a different subsystem of society (e.g., the market *j*). The corresponding equation can be formulated as follows:

$$x_{t+1}^i = ax_t^i(1 - x_{t+1}^j) \qquad (3)$$

These transversal selections among subsystems are additional to the longitudinal development of meaning at all levels within self-referential systems. Note that the longitudinal selection is hierarchical, while the transversal one changes with the nature of the differentiation. The two axes stand orthogonally under the condition of functional differentiation. In other words, the longitudinal generation of meaning was already available as an incursive mechanism in pre-modern societies, while the transversal updates provide modern societies with an analytically independent mechanism for organizing communications.

## 4. Functional differentiation

Let us return to Equation 1 to understand what functional differentiation means for a system that can operate both incursively (that is, against the axis of time) and recursively (that is, historically). Adding the time subscripts, one can first reformulate Equation 1 for the (more general) recursive case:

$$S_{t+1} = f(S_t, E_t) \tag{5}$$

When functionality of the differentiation prevails, the external environment *E* is decomposed for each subsystem in other subsystems (with a remaining term ε as a representation of the residual environment). This can modeled at the subsystems level—let me use the lower-case *s* for this level—as follows:

$$s^i_{t+1} = f(s^i_t, s^j_t, s^k_t, s^l_t, ..., \varepsilon_t) \tag{6}$$

In a functionally differentiated system, the windowing of the subsystems upon each other becomes horizontal. A biological system, however, would remain also integrated also for the sake of survival. A communication system can also develop in terms of differentiated fluxes that are integrated by being organized historically.

How might functional differentiation in the communication take control over development away from the historical organization of a social system? Luhmann (1997, 2000) distinguished between the self-organization of the differentiated fluxes of communication and their integration into organizations. The organization of interfaces is historical, while the self-organization of the differentiated communication operates evolutionarily, that is, by providing meaning to the events with hindsight and by using potentially different codes. The social system would thus be able to entertain a set of models of itself. In other words, an additional equation can be formulated for each subsystem given selections by the others from the perspective of hindsight:

$$s_{t+1}^i = f(s_t^i, s_{t+1}^j, s_{t+1}^k, s_{t+1}^l, \ldots, \varepsilon_t) \qquad (7)$$

Note that all systems and subsystems continue to operate historically, and thus provide meaning to their own development along the longitudinal axis at the same time as they provide meanings to one another. The relations between subsystems can be expected to function to variable degrees both incursively and recursively because Equation 6 and Equation 7 are both operational. However, the functionally differentiated system contains an additional Δ$t$ at each interface within the system. This Δ$t$ can be used for a local reversal of the time axis and thus generate a transversal incursion which stands orthogonally to the longitudinal incursions.

The two types of incursion—the transversal and the longitudinal ones—stand orthogonally, but they can interact in systems that are able to *process* meaning. This interaction can perhaps be considered as the crucial difference between biological and social systems. Differentiated biological systems can hold different meanings, but are not supposed to vary them over time for reasons other than natural one (that is, survival purposes). The integrity of these systems would be endangered if the codes were varied, that is, by entertaining them playfully. Social systems, however, allow for this type of (potentially innovative!) processing and recombination of meanings. Interactions among differently codified meanings over time provide us with other combinations in a horizon of possible meanings.

Using the logistic equation, one can formulate Equation 7 for the case where other subsystems provide relevant environments for the development of subsystem *i*, as follows:

$$x_{t+1}^i = ax_t^i(1 - x_{t+1}^j)(1 - x_{t+1}^k)(1 - x_{t+1}^l)... \; \varepsilon \qquad (8)$$

Each subsystem (*i*) develops with reference to its own previous state, but one can expect that all other subsystems feedback upon this development by entertaining a model of the reference system in the present using their respective codes. Since each meaning-providing subsystem (*i*) also provides meaning to its own development longitudinally, and the two types of meaning-providing can interact, one can generalize Equation 8 as follows:

$$x_{t+1}^i = ax_t^i \prod_{1}^{n}(1 - x_{t+1}^n).\varepsilon \qquad (9)$$

In this formula *n* represents the number of subsystems of the functionally differentiated system. While this number was analytically restricted in Parsons's structural-functionalism—using his so-called four-function paradigm—the number of subsystems can vary in Luhmann's (1997) theory with the historical development of the media of communication and their symbolic generalization into codes. When the subsystems use different frequencies for their updates, parameters have to be added to the corresponding selection mechanisms.

**5. Stabilization, Meta-stabilization, and Globalization**

Let us now turn to the question of what one can expect when one type of meaning operates upon a differently codified meaning at one or more interfaces. I mentioned above that the operation of two differently codified incursions upon each other may lead to "mutual shaping" and the consequent stabilization of a co-evolution along a trajectory. The formalization will enable us to distinguish between stabilization in a co-evolution between two subdynamics and the possibility of globalization in the case of three subdynamics.

When selection is represented by the feedback term of the logistic equation, that is, by $(1 - x)$, two selections of the same system operating on a variation $a$ would result at the systems level in a quadratic expression of the following form:

$$f(x) = a (1 - x)(1 - x)$$
$$= a (x^2 - 2x + 1) \qquad (10)$$

This function is represented by the solid line in Figure 2a (on the left side): a system with two selections can be stabilized at the minimum of a quadratic curve. When this minimum is extended along the time dimension, a valley is shaped in which the system can follow a trajectory (Sahal, 1985; Waddington, 1957).

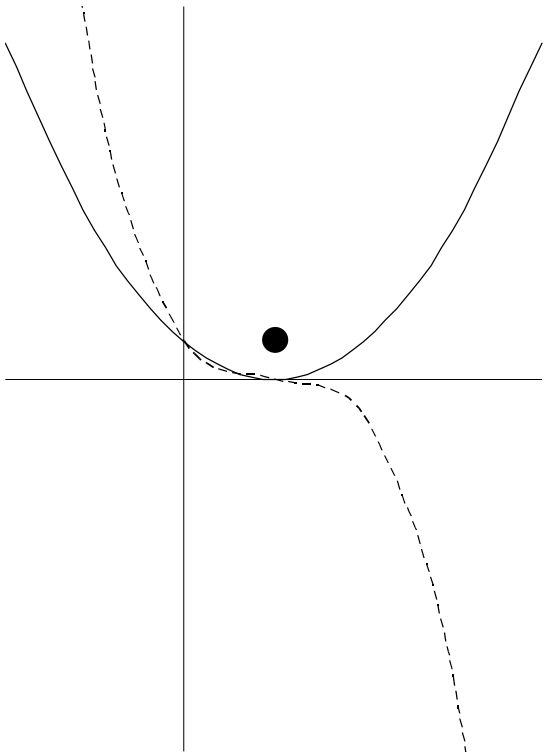 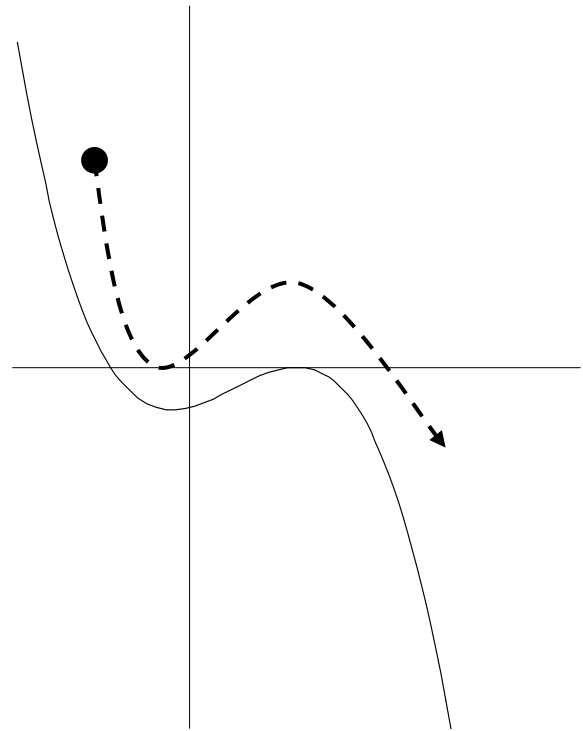

**Figure 2a**:
— $f(x) = x^2 - 2x + 1$ (stabilized)
- - - $f(x) = x^3 - 3x^2 + 3x - 1$ (globalized)

**Figure 2b**: Globalization and stabilization are no longer harmonized; a local optimum can be sustained temporarily.

Adding one more selection term leads analogously to the following equation:

$$f(x) = a\,(1-x)(1-x)(1-x)$$
$$= -a\,(x^3 - 3x^2 + 3x - 1) \qquad (11)$$

This latter function is represented as the dotted line in Figure 2a. As long as the selections operate with the same parameters, the global and the stable points of inflection coincide (in a so-called "saddle point"). The historically stabilized system can then be identical with the global one. Perhaps such a harmonized system can be said to have an identity, since the global optimum coincides with the localized one.

Figure 2b shows the resulting configuration when stabilization and globalization no longer operate with the same parameter values. In this (more general) case, one would expect the curve to show both a maximum and a minimum. At the minimum the system is stabilized, but at the maximum it can be considered as meta-stable. A bifurcation is thus induced because the system can go backward (to the stabilization of an identity) or forward (to globalization into a next-order regime). As long as the system remains stable (that is, at the minimum), it can develop along a trajectory. However, the flux tends to move the system towards the other basin of attraction. This attraction is caused by the possibility to communicate in an additional dimension, and thus to process more complexity in the newly emerging configuration (Turing, 1942).

The sign of the equations merits further attention. Equation 10 has a positive sign if one assumes the logistic equation as the basic format for the selection. If this sign is reversed, another subdynamic has to play a role because this inversion cannot be endogenous to the mutual selections that co-evolutionarily stabilized the system. The third (sub)dynamic may either reinforce the prevailing equilibrium and make the system hyper-stable, or invert the sign and make the system meta-stable (Figure 3). Thus, both meta-stability and hyper-stability indicate that a third subdynamic is operating (but not yet necessarily manifest).

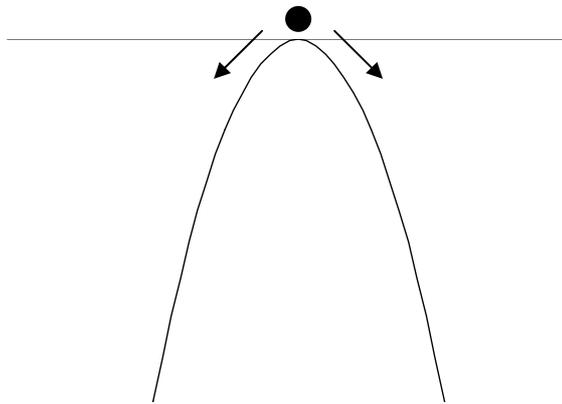 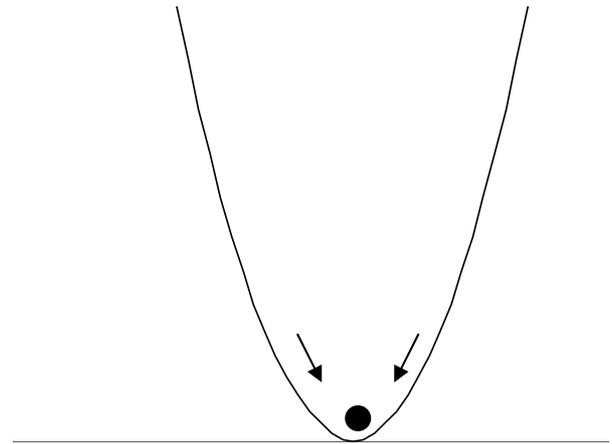

**Figure 3a**: Meta-stabiliziation           **Figure 3b:** Hyper-stabilization

For example, the change in the sign can be appreciated as a market with decreasing marginal returns versus one with increasing marginal returns, as in the case of information and communication technologies (Arthur, 1988, 1989). Increasing marginal returns can lead to a bifurcation at the meta-stable vertex of the hyperbole and a consequential lock-in on either side. In the case of a lock-in, the previously meta-stable system is hyper-stabilized by using the third term as a feedback which reinforces the co-evolution between the two other subdynamics (Leydesdorff & Meyer, forthcoming).

Figure 2b above provided the full picture with both a positive and a negative inflection point in the case of three interacting subdynamics. Stabilization can be considered as a result of integration (e.g., by organization), while differentiation among the self-organizing fluxes can be expected to prevail when the system is less organized. However, the distinction between these two subdynamics—integration and differentiation—remains analytical; in the social system they can be expected to concur, since the global system cannot be historically manifest without some form of stabilization occurring at the same time. The interactions between the various subdynamics make the system complex and result in the expectation of continuous transitions between provisional (that is, local) stabilizations versus globalizations at the systems level. How would these subdynamics of stabilization and globalization operate in the case of a meaning-processing system?

**6. Hyper-incursion**

The historical realization of meaningful information by the first incursion at each moment (Selection 1 in Figure 1) remains a necessary condition for the development of the social system, but this operation does not have to be attended by the other two selection mechanisms continuously. The other two selection mechanisms can also be expected to interact. A third incursive mechanism can interact with the interaction term between two incursive selections, and this can generate a next-order selection or hyper-incursivity. A hyper-incursive equation no longer refers necessarily to a historical realization because it can interact with the interaction term between the two anticipatory mechanisms which refer to the new state.

The most radical of the hyper-incursive equations reflects this orientation to the future as follows:

$$x_t = a x_{t+1} (1 - x_{t+1}) \qquad (12)$$

This system no longer contains any reference to its previous state $x_{t-1}$ or its current state $x_t$ (at $t = t$), but the emerging state is considered as a function of different expectations about the future. Note that both expectations interfaced in this equation can only be based on previous incursions (Figure 4). When incursions are interfaced *recursively*, only a historical variation can be produced (Leydesdorff, in print). However, when the two incursions are interfaced *incursively*, hyper-incursion can also be expected.

*Selection 3:*
Hyper-incursion with two references to $t = t + 1$

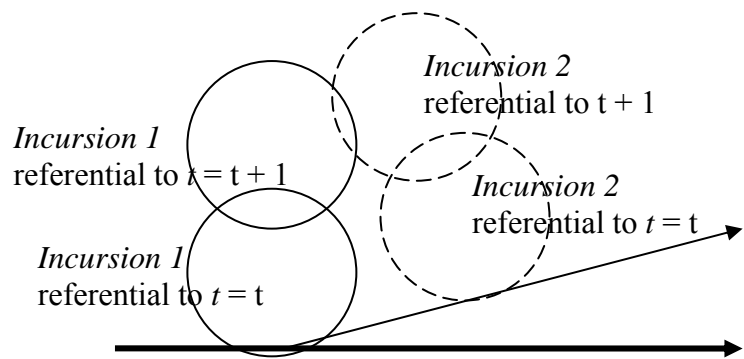

**Figure 4**: Hyper-incursivity at the interface between two incursive selections: *Incursion 1* as selection on the variation at each moment of time, and *Incursion 2* as stabilization with reference to change over time.

While each incursion generates meaningful information and thus becomes organized both historically (with reference to $t = t$) and reflexively (with reference to $t = (t + 1)$), hyper-incursion is by itself not yet organized at $t = t$ because it contains only a reference to $t = (t + 1)$. An additional subdynamic, therefore, would be needed in order to make the results of this hyper-incursion again organized in historical time. Luhmann (2000) hypothesized that self-organization among the (differently coded) fluxes of communication—which generate only expectations—is brought under organizational control by making decisions. Let us first see how this works in terms of the equations.

Equation 12 can be rewritten as follows:

$$x_t = a x_{t+1} (1 - x_{t+1}) \qquad (12)$$

$$x_t = a x_{t+1} - a x_{t+1}^2$$

$$a x_{t+1}^2 - a x_{t+1} + x_t = 0$$

$$x_{t+1}^2 - x_{t+1} + (x_t/a) = 0 \qquad (13)$$

For $a = 4$, $x_{t+1}$ is defined as a function of $x_t$ as follows:

$$x_{t+1} = \tfrac{1}{2} \pm \tfrac{1}{2} \sqrt{(1 - x_t)} \qquad (14)$$

Depending on the plus or the minus sign in the equation, two future states are generated at each time step. Since this formula is iterative, the number of future states doubles with each next time step. After $N$ time steps, $2^N$ future states would be possible. (For $N = 10$, the number of options is already more than a thousand.)

Dubois (2003, at p. 115) specified a decision function u(t) that can be added for making a choice between two options:

$$u(t) = 2\, d(t) - 1 \qquad (15)$$

where $u = +1$ for the decision $d = 1$ (true) and $u = -1$ for the decision $d = 0$ (false). In a social system, however, more choices than these two extremes are possible. Social systems operate in a distributed mode, and therefore one would expect a probability distribution of preferences. This distribution contains an uncertainty. In distributed systems, decisions can be organized and codified into decision rules (Bertsekas & Tsitsiklis, 1989).

Note that the decisions or decision rules do not determine the hyper-incursive dynamics of the system, but only guide these dynamics historically. (A codified decision rule could function as yet another incursive mechanism.) However, hyper-incursion continues to create possible futures as interactions among expectations on the basis of incursive subdynamics. Decisions change the *historical* conditions by organizing the system, that is, by closing the circle in Figure 1. However, decisions don't have to be taken at each step. Without historicity, the interfacing of expectations would remain in the virtual realm of fantasy and speculation.

The horizon of possible meanings itself could not be further developed historically without any social realizations. The relevant decisions, however, are no longer taken about historical facts, but about expectations. Thanks to this orientation towards the future, the social system can become increasingly knowledge-based because anticipations are continuously being fed into its historical development. The historical realizations *result* increasingly from the interfacing of expectations as a base other than—but in interaction with—the historical organization of "reality" (Equation 6). The cycling among the selections adds meaningful information to the information flux which develops with the time axis. By bending the time-axis, the meaning-generating system increasingly reconstructs its own history (Hellsten *et al.*, 2006).

In other words, decisions specify instantiations among the available options when anticipations are interfaced. The organization of decisions along a historical axis potentially stabilizes a trajectory within the phase space of expectations. Social order is made contingent on coupling expectations about other possible realities with the historical reality of observable actions and institutions. The latter are increasingly reconstructed as the results of interactions among expectations.

Historically, one may be inclined to consider actions and institutions as the causes of expectations because the former precede the latter. However, anticipations are based on the perspective of hindsight, while the latter becomes analytically prior to the historical sequence of events if the time axis is locally inverted. When the feedback is codified and the codification reinforced by global horizons of meaning, the reconstructive arrow (from the perspective of hindsight) can in the longer term become more important than the forward arrow of historical development. While the historical construction continues to feed into the cycle bottom-up, control in communication systems thus tends to become top-down when the cycle can be closed and reinforced as a next-order dynamics. Unlike "natural" preferences grounded in the historical direction of the time axis, informed decisions guide development into an increasingly knowledge-based order given the contingencies of historical situations as relevant contexts.

**7. Mutual anticipation in social systems**

Two other hyper-incursive models follow from the logistic equation:

$$x_t = ax_t(1 - x_{t+1}) \qquad (16)$$

$$x_t = ax_{t+1}(1 - x_t) \qquad (17)$$

- Equation 16 evolves into $x = (a - 1)/a =$ Constant. I submit that this evolution towards a constant through anticipation can be considered as modeling the self-reference or the expectation of an (individual) identity. Identity is based on the expectation of continuity of the "self" in the next stage. At the level of the social system a group can also be expected to have an identity;
- Equation 17 evolves into $x_{t+1} = (1/a) \{x_t / (1 - x_t)\}$. Since the latter term approaches $-1$ as its limit value and the former term is a constant $(1/a)$, this representation can alternate between itself and a mirror image. This subdynamic thus formalizes the reflexive operation.

In other words, the three hyper-incursive equations specify the fundamental processes of meaning-processing: identity formation (Eq. 16), reflexivity (Eq. 17), and the mutual expectation of expectations (Eq. 13). While identity and reflexivity are more commonly defined in social theory, the expectation of expectations among systems may require additional explanation. Let me develop this as an analogon of the concept of "double contingency," but at the level of differently coded subsystems of society.

"Double contingency" was theorized in sociology as an interaction between *Ego* and *Alter* (Parsons, 1968). *Ego* and *Alter* are used in order to emphasize that in the relation between two human beings, the *Ego* knows the *Alter* to be a reflexive *Cogito* who entertains expectations. Thus, a social reality is constructed in which a symbolic order is invoked (Lévy-Strauss, 1987; Elmer, 1995). For the *Ego* this means that one's own reflexivity is reinforced in the encounter. The *Ego* in the present ($x_t$) no longer has a reference to itself other than in a future state ($x_{t+1}$), that is, as an *Alter Ego*. The orientation of providing meaning with reference to other possible meanings thus constitutes the social world as an order different from psychological ones (Husserl, 1929; Schutz, 1932).

Habermas (1987) criticized Luhmann's social systems theory for having replaced the traditional subject of transcendental philosophy (that is, the *Ego*) with the concept of a social system. Let me use this critique as a heuristics for understanding meaning-processing in a social system. Unlike the multitude of subjects—who can entertain relations and experience double contingency in these relations—there is only a single social system. This system can contain Luhmann-type subsystems at each moment and Rosen-type models of itself over time. One can expect differently codified *sub*systems of communication only under the condition of functional differentiation.

These different subsystems interact in historical organizations, but additionally they entertain and develop expectations which contain references to other possible expectations. Given this feedback, the reference of an organization to the time horizon may evolve from developing within history to developing in terms of expectations about the future, and thus to being increasingly knowledge-based. Note that the phenotypical organizations remain a mixture of the two principles of organization, since both Equations 6 and 7 can be expected to operate, albeit with variable weights.

Unlike the self-organizing fluxes of communication in social systems, organized systems contain a reference to themselves as an identity (Equation 16). (Individuals can then be considered as the smallest units which organize the fluxes.) The social order of expectations remains structurally coupled to historical formats of organization because the systems dynamics cannot otherwise become historical. The self-organization of fluxes in terms of functional differentiations is not itself historically observable, but only perceptible after a reflexive turn (Husserl, 1929; Luhmann, 1984). The coupling between the social system of communications and the agents carrying it is provided by the reflexivity formalized in Equation 17. Because of this reflexivity, the social system can be entertained by individuals as a notional order, that is, an order with the epistemological status of a hypothesis.

*Vice versa*, the individuals and institutional agents are reflected in the social system as addresses or nodes. Since both systems can be expected to contain the reflexive operation expressed in Equation 17, a coevolution ("mutual shaping") can also take place between the social system and individuals. However, this coevolution can be interrupted when expectations are further codified (e.g., stabilized) on either side beyond the control of the other side. On the side of social systems, an eigendynamics of codifications may lead to "alienation" (Platt & Weinstein, 1971). On the side of individuals, the recursive application of the reflexive operation provides discretionary room for private thinking and tacit knowledge.

In summary, functional differentiation changes the social system so that a double contingency between functionally differentiated (that is, differently coded) subsystems of communication becomes possible. Each of the subsystems can be expected to generate its own set of expectations. As these sets of expectations are made relevant to one another, their identity is no longer given, but has to be organized. Organization means in this case limiting the range of options by making decisions about the relevant expectations. For example, in the case of innovation one combines market perspectives with technological options. From this perspective, the individual (e.g., the entrepreneur) can be considered as the minimal carrier of decisions because decision-making itself can also be a social process. As was shown, this process guides the self-organizing dynamics by organizing it, since the complexity would otherwise explode. In technical terms, development would rapidly become non-computable because it would be non-polynomially (NP) complete (Equation 14).

**8. Conclusions**

Like all self-referential systems, social systems can generate meaning in an incursive mode and thus be anticipatory, that is, by entertaining expectations along the axis of time. The functional differentiation of society in terms of different codes of communication provides the social system with an additional mechanism of anticipation because of asynchronicity in the updates among the different subsystems. In organized systems and hierarchies, these transversal updates at subsystemic levels are synchronized at the top with reference to an external environment, but a functionally differentiated system has replaced this option for survival by considering the subsystems as relevant environments for one another.

The transversal generation of meaning among the subsystems can be recombined with the longitudinal generation of meaning and thus provide a field of possible meanings. This second-order incursion upon the anticipatory terms of the first-order incursions provided us with the hyper-incursive equation which models a third selection mechanism. The uncertainty generated by this model, however, forces the system to be organized periodically like an update mechanism. I submitted that the minimum unit for this historicity is the individual.

At the level of individual interactions, hyper-incursivity is also available as a "double contingency": the meeting of the other as entertaining a similar, but potentially different set of expectations. "Double contingency" can thus be considered as a transversal encounter between different sets of meanings, but in the case of individuals whole systems (agents) meet and update, and not subsystems. Subsystems of communication operate at a different level, and they can only reproduce a double contingency towards one another if the functional differentiation between them is warranted in terms of codes developing along different axes.

The social system shares with psychological systems the other two hyper-incursive operations (identity formation and reflexivity). Unlike individuals, the social system is able to generate within the system a non-linear dynamic among *three* selection mechanisms: (1) selections from a *variety* of global horizons of meaning, (2) historical realizations in organizations at each moment, and (3) the transformation of the latter by the need to recombine expectations based on previous realizations over time. Individuals can meet and experience a double contingency in the relation between *Alter* and *Ego*, but by doing so they already constitute a social system (Leydesdorff, 2003; Leydesdorff & Dubois, 2004).

Thus, a hypercycle between three selection mechanims of meaning can be shaped at the level of the social system. The three selections operate as in a triple helix: when the configuration allows for interaction of a third selection mechanism with the interaction between two other selections, hyper-incursion becomes a possibility, and when this occurs the systems can be expected to avalanche towards the strongly anticipatory regime of a knowledge-based society. In a knowledge-based order, decisions can increasingly be informed by expectations.

Let me finally note that the argument in this paper would indicate that the social system as a strongly anticipatory one would *not* be able to construct its own next stage in the longer run without historical decision-making by an external agency. Thus, social systems can be considered as semi-autonomous (Collier, 2005). While Dubois (2002) showed that a strongly anticipatory system is able to predict the development of a complex system given a fixed relation between the modeling and the modeled system, this relation can be expected to remain variable in social systems because the modeling system provides meaning to the modeled system using a non-linear dynamics. Decision-making structures (e.g., individuals) are a necessary condition for the historical development of the non-linear dynamics of meaning in social systems.

**Acknowledgment**

I thank John Collier for his comments on an earlier draft.